\documentclass[conference]{IEEEtran}
\IEEEoverridecommandlockouts
\usepackage{cite}
\usepackage{amsmath,amssymb,amsfonts}
\usepackage{graphicx}
\usepackage{textcomp}
\usepackage{comment}
\usepackage{float} 

\usepackage{tikz}
\usetikzlibrary{arrows.meta,positioning,shapes.geometric,calc,fit, shadows, backgrounds}

\tikzstyle{block} = [rounded rectangle, minimum width=1cm, minimum height=1cm, text centered, draw=black, thin, inner sep=2pt]
\tikzstyle{arrow} = [thick,->,>=stealth]

\usepackage{color}

\usepackage[utf8]{inputenc} 
\usepackage[T1]{fontenc}    
\usepackage{hyperref}       
\usepackage{url}            
\usepackage{booktabs}       
\usepackage{amsfonts}       
\usepackage{nicefrac}       
\usepackage{microtype}      
\usepackage{xcolor}         
\usepackage{enumitem}
\usepackage{comment}
\usepackage{tcolorbox}
\usepackage{listings}
\usepackage{pdfpages}
\usepackage{longtable}
\usepackage{multirow}
\usepackage{graphicx}
\usepackage{verbatim}
\usepackage{stfloats}
\usepackage{balance}
\usepackage[export]{adjustbox}

\usepackage{tabularx}
\usepackage{lipsum} 
\usepackage{colortbl}
\usepackage{float}
\usepackage{hyphenat}

\usepackage{booktabs}
\usepackage{array}
\usepackage{wrapfig}
\usepackage{fancyvrb}

\usepackage{setspace}
\definecolor{swecream}{HTML}{EFFEFF}
\definecolor{issueborder}{HTML}{15071A}
\definecolor{issuefill}{HTML}{F6F8FA}
\definecolor{envfill}{HTML}{F2F9FF}
\definecolor{envborder}{HTML}{123C7C}
\definecolor{agentfill}{HTML}{F9F3F3}
\definecolor{agentborder}{HTML}{9B0A0A}
\definecolor{goldpatchborder}{HTML}{FABB00}
\definecolor{goldpatchfill}{HTML}{FFF7E1}

\DefineVerbatimEnvironment{CodeVerbatim}{Verbatim}{
  formatcom={\color{black}},
  fontsize=\small,
  fontfamily=\ttdefault,
  fontseries=\mddefault,
  fontshape=\updefault,
  fillcolor=\color{white},
  framerule=0pt,
}
\usepackage{longtable}
\newtcolorbox{observationbox}[1][]{
        colback=envfill,
        colbacktitle=envfill,
        colframe=envborder,
        arc=5pt,
        fontupper=\small,
        fonttitle=\bfseries\color{black},
        boxrule=0.5mm,
        boxsep=1mm,
        width=\linewidth,
        breakable,
        title={Observation \hfill #1},
        rounded corners,
        toptitle=0.7mm,
        bottomtitle=0.7mm
}
\newtcolorbox{goldpatchbox}[1][]{
        colback=goldpatchfill,
        colbacktitle=goldpatchfill,
        colframe=goldpatchborder,
        arc=5pt,
        fontupper=\small,
        fonttitle=\bfseries\color{black},
        boxrule=0.5mm,
        boxsep=1mm,
        width=\linewidth,
        breakable,
        title={\twemoji{1f6a9} Flag Captured \hfill #1},
        rounded corners,
        toptitle=0.7mm,
        bottomtitle=0.7mm
}
\newtcolorbox{issuebox}[1][]{
        colback=issuefill,
        colbacktitle=issuefill,
        colframe=issueborder,
        arc=5pt,
        fontupper=\small,
        fonttitle=\bfseries\color{black},
        boxrule=0.5mm,
        boxsep=1mm,
        width=\linewidth,
        breakable,
        title={Issue \hfill #1},
        rounded corners,
        toptitle=1mm
}
\newtcolorbox{agentbox}[1][]{
        colback=agentfill,
        colbacktitle=agentfill,
        colframe=agentborder,
        arc=5pt,
        fontupper=\small,
        fonttitle=\bfseries\color{black},
        boxrule=0.5mm,
        boxsep=1mm,
        width=\linewidth,
        breakable,
        title={EnIGMA \hfill #1},
        rounded corners,
        toptitle=1mm,
        lower separated=false
}
\newtcolorbox{fileviewerbox}[1]{
        enhanced,
        breakable,
        boxrule = 1.5pt,
        fontupper = \small,
        fonttitle = \bf\color{black},
        arc = 5pt,
        rounded corners,
        colframe = black,
        colbacktitle = swecream,
        colback = swecream,
        title = #1,
        left=4pt 
}
\newtcolorbox{promptbox}[1]{
    enhanced,
    breakable,
    boxrule=1pt,  
    fontupper=\small,
    fonttitle=\bfseries\color{black},
    arc=3pt,  
    rounded corners,
    colframe=black,
    colbacktitle=swecream,
    colback=swecream,
    title=#1,
    left=2mm,  
    right=2mm,  
    top=1mm,  
    bottom=1mm  
}

\newfloat{algorithm}{tbp}{loa}
\floatname{algorithm}{Algorithm}
\newcounter{algline}
\newenvironment{algorithmic}[1][]{%
  \begin{list}{\arabic{algline}:}{%
    \usecounter{algline}%
    \setlength{\leftmargin}{1.3em}%
    \setlength{\labelwidth}{1.1em}%
    \setlength{\labelsep}{0.2em}%
    \setlength{\itemsep}{0pt}%
    \setlength{\parsep}{0pt}%
    \setlength{\topsep}{1pt}%
  }%
}{\end{list}}
\newcommand{\STATE}{\item}
\newcommand{\IF}[1]{\item \textbf{if} #1 \textbf{then}}
\newcommand{\ELSE}{\item \textbf{else}}
\newcommand{\ENDIF}{\item \textbf{end if}}
\newcommand{\RETURN}{\item \textbf{return} }

\newcommand{\cmark}{\textcolor{green!60!black}{$\surd$}}
\newcommand{\xmark}{\textcolor{red}{$\times$}} 
\newcommand{\pmark}{\textcolor{orange!85!black}{$\triangle$}}
\newcommand{\rev}[1]{#1}

\makeatletter

\def\BibTeX{{\rm B\kern-.05em{\sc i\kern-.025em b}\kern-.08em
    T\kern-.1667em\lower.7ex\hbox{E}\kern-.125emX}}
\begin{document}


\title{ThinkNet: Evolutionary Deep Ensembles for Efficient Motor-Imagery EEG Classification}

\author{
\IEEEauthorblockN{Abdul Basit, Saim Rehman, Muhammad Shafique}
\IEEEauthorblockA{\textit{eBRAIN Lab, Division of Engineering} \textit{New York University (NYU) Abu Dhabi}, Abu Dhabi, UAE\\
\{abdul.basit, sr7849, muhammad.shafique\}@nyu.edu}
\vspace{-20pt}
}


\maketitle

\begin{abstract}
\rev{Subject-independent }motor-imagery EEG (MI-EEG) \rev{decoding requires models }that generalize to \rev{unseen users when }target-user data \rev{are unavailable during model and policy selection, while supporting resource-constrained inference}. However, held-out-subject performance can be overstated when test-subject information influences preprocessing, model selection, or ensemble selection. We present \textit{ThinkNet}, a validation-controlled framework that combines train-only normalization, validation-guided evolutionary search, and validation-gated inference to \rev{select }compact decoders and inference policies for held-out subjects. We evaluate four-class BCI Competition IV-2a (session T) decoding with nine LOSO folds, three seeds, seven fixed entries, and ten representative decoder families\rev{, plus a frozen-configuration LOSO check on 54-subject Lee2019 dataset. Fold-specific search, normalization, and policy selection use no held-out-subject feedback. The seven-entry comparison separately evaluates frozen post-discovery configurations}. In the \rev{broad family-selection benchmark, source validation exceeded the uniform-choice expectation by 1.72 points but did not outperform a fixed-family EEGNet policy (40.01\% vs. 40.29\%), remained 6.08 points below the test-best family, and showed weak within-fold rank transfer ($\rho=0.256$). Restricting eligibility to models with $\leq$25K parameters produced an observed mean difference of $-0.05$ points. Among five frozen configurations, Lee2019 provided an independent consistency check for the size--performance and selector-regret patterns. Separately, the fixed }compact decoder achieved 44.35$\pm$\rev{15.70\% }with 4.9K parameters, 19 \rev{KiB }FP32 weights, and 0.99 ms batch-1 Orin CUDA inference. \rev{Compact-six validation gating improved mean accuracy by 2.47 points (40.84\% to 43.31\%) at higher measured inference cost. ThinkNet therefore makes source-only compact model }and \rev{policy selection explicit and auditable, while exposing selector reliability as the principal limitation under subject shift}.
\end{abstract}


\section{Introduction}
\label{sec:intro}
Motor-imagery brain--computer interfaces (MI-BCIs) translate imagined movement, such as left hand, right hand, feet, or tongue imagery, into control commands from non-invasive EEG signals. This capability is a key step toward assistive communication, neurorehabilitation, and closed-loop prosthetic control, where the decoder must operate reliably from noisy, low-amplitude sensorimotor rhythms~\cite{nicolas2012brain,daly2008brain}. Despite steady progress, practical MI decoding remains constrained by three persistent issues: large inter-subject variability, session non-stationarity in longitudinal use, and the need for low-latency inference on resource-limited devices~\cite{tangermann2012review,lotte2018review}. A method that performs well after subject-specific tuning may therefore be much less useful when deployed on a previously unseen user.

Deep learning has improved EEG representation learning by reducing reliance on fully hand-crafted pipelines and learning temporal, spectral, and spatial filters directly from data. EEGNet introduced compact depthwise and separable convolutions tailored to EEG~\cite{lawhern2018eegnet}. However, reported accuracies across the literature often depend on different preprocessing choices, subject/session splits, model capacities, augmentation strategies, and selection rules, especially when validation decisions are not clearly separated from held-out testing. This makes it difficult to answer a practical question: \textit{which compact decoder and inference policy should be selected when the target user is unavailable during training, and how reliable is validation-based selection under subject shift?}

A motivating case study from our search results illustrates the \rev{source-only model-selection }challenge. Source-validation family selection reached 40.01\% held-out accuracy, \rev{versus 38.29\% for uniform choice and 40.29\% for fixed-family EEGNet, }whereas a non-deployable \rev{test-best family }reached 46.09\%\rev{; within-fold rank transfer was weak ($\rho=0.256$). Thus, when a }new user is unavailable for tuning, validation \rev{contains measurable but insufficient information for reliable family selection. ThinkNet makes that limitation explicit and auditable.}

\textit{ThinkNet is built around this subject-independent selection problem.} 
Unlike target-adaptive transfer-learning methods, it studies zero-target model and inference-policy selection under held-out-subject evaluation. We use strict leave-one-subject-out (LOSO) evaluation on BCI Competition IV-2a (session T), separate source-subject validation from held-out-subject testing, search compact and broader decoder-family configurations, and report accuracy\rev{, size, }and selection behavior. \rev{Fold-specific search and gating use }train-only normalization \rev{and no held-out-subject selection feedback. The fixed benchmark is reported separately as a post-discovery comparison}.
\rev{ThinkNet separates two decisions: whether a fixed compact default is sufficient, and whether source evidence supports a different model or higher-cost policy.} Thus, ThinkNet is a validation-controlled selection and evaluation protocol rather than a new convolutional block. \rev{Evolutionary search, compact decoders, and ensembles are not individually claimed as new; the contribution is their integration with target isolation, auditable decision provenance, and explicit selector-regret measurement.}
\underline{\textbf{Novel Contributions:}}
ThinkNet's main contributions are:
\begin{enumerate}[leftmargin=*, itemsep=1pt, topsep=2pt]
    \item \textbf{A validation-controlled ThinkNet framework} for compact subject-independent MI-EEG decoder and inference-policy selection, independent of any single architecture family.
    \item \textbf{A \rev{selector-reliability analysis}} \rev{that measures within-fold rank transfer, uniform}/\rev{fixed-family}/\rev{test-best references, and family-level regret with subject-bootstrap uncertainty and an independent Lee2019 consistency check}.
    \item \textbf{A representative ten-family \rev{accuracy--size }search} \rev{with }validation-selected %
\rev{retraining (}15,390 \rev{candidates; }270 \rev{checkpoints) and a budget-matched random-search control that separates compactness from held-out accuracy gains}.
    \item \textbf{A validation-gated \rev{inference-policy analysis}} \rev{comparing }single-model inference, averaging, and voting using validation \rev{only, with measured Orin accuracy--cost operating points}.
    \item \textbf{A \rev{strict, reproducible LOSO protocol}} \rev{with train-only normalization, explicit split accounting, three-seed evaluation, and auditable decision provenance}.
\end{enumerate}
Together, these contributions position ThinkNet as an auditable accuracy--efficiency model-selection framework for subject-independent MI-EEG decoding, rather than a single-number benchmark claim.

\section{Related Work}
\label{sec:related}
Classical MI-BCI pipelines commonly combine band-pass filtering, spatial filtering, and shallow classifiers. Filter-bank common spatial patterns and Riemannian covariance methods remain influential because they encode neurophysiological structure and can remain effective with limited data~\cite{4634130,6046114}. These methods also established the need for careful validation, since small EEG datasets are highly sensitive to subject identity, session drift, and preprocessing choices~\cite{lotte2018review}. However, they typically depend on hand-designed feature choices, and their strict held-out-subject model-selection behavior is less often analyzed as the main object of study.

Deep MI decoders learn temporal, spectral, and spatial representations directly from EEG trials. EEGNet is especially relevant for embedded BCI because depthwise and separable convolutions keep the parameter count small~\cite{lawhern2018eegnet}; Shallow ConvNet provides a band-power-inspired convolutional baseline~\cite{schirrmeister2017deep}. EEG-TCNet and EEG-Inception extend this line with temporal convolution and multi-scale branches~\cite{ingolfsson2020eeg_tcnet,zhang2021eeg_inception}. FBCNet preserves filter-bank priors within a learned model~\cite{mane2021fbcnet}, while ATCNet, EEG-Conformer, and compact convolutional transformers add attention or conformer-style temporal modeling~\cite{altaheri2022atcnet,song2023eegconformer,keutayeva2024eegcct}. These architectures broaden the design space, but reported performance still depends strongly on preprocessing, augmentation, subject/session splits, and whether validation choices are separated from held-out testing.

\begin{table}[ht]
\centering
\caption{Positioning of ThinkNet relative to MI-EEG literature.}
\label{tab:related_positioning}
\scriptsize
\setlength{\tabcolsep}{1.5pt}
\renewcommand{\arraystretch}{1.04}
\resizebox{\linewidth}{!}{
\begin{tabular}{>{\raggedright\arraybackslash}p{0.47\linewidth}cccc}
\toprule
\textbf{Method family} & \textbf{Compact} & \textbf{SI/LOSO} & \textbf{Search/sel.} & \textbf{Sel.-gap} \\
\midrule
CSP/FBCSP, Riemannian~\cite{4634130,6046114} & \cmark & \pmark & \xmark & \xmark \\
Compact/temporal CNNs~\cite{lawhern2018eegnet,ingolfsson2020eeg_tcnet} & \cmark & \pmark & \xmark & \xmark \\
Filter-bank/attention/conformer models~\cite{mane2021fbcnet,altaheri2022atcnet,song2023eegconformer,keutayeva2024eegcct} & \pmark & \pmark & \xmark & \xmark \\
Transfer/domain adaptation~\cite{lotte2018review,wu2020transfer,wei2021intersubject} & \pmark & \cmark & \pmark & \xmark \\
EEG NAS / general ensembles~\cite{JMLR:v20:18-598,duan2022ctnaseeg,lakshminarayanan2017simple} & \pmark & \pmark & \cmark & \pmark \\
\textbf{ThinkNet} & \cmark & \cmark & \cmark & \cmark \\
\bottomrule
\end{tabular}}
\parbox{\linewidth}{\vspace{2pt} \scriptsize\emph{Note:} \cmark{} denotes central focus, \pmark{} partial/protocol-dependent coverage, and \xmark{} not primary. SI/LOSO denotes subject-independent or leave-one-subject-out evaluation; Search/sel. denotes automated architecture, family, hyperparameter, or \rev{policy }selection; \rev{Sel}.\rev{-gap }denotes \rev{explicit auditing of source-validation choices against isolated held-out-subject outcomes}.}
\end{table}

Subject-independent and transfer-learning methods address inter-subject variability by learning source-invariant representations, adapting across users, or fine-tuning with available target-domain information~\cite{lotte2018review,wu2020transfer,wei2021intersubject}. This is complementary to ThinkNet: our focus is validation-controlled selection when the held-out subject is unavailable for normalization fitting, architecture selection, hyperparameter choice, or ensemble-policy selection. \rev{Methods using unlabeled target epochs or labeled calibration trials address a different target-access setting; ThinkNet evaluates the source-only stage that could precede such adaptation.} EEG architecture search and the broader deep-ensemble literature offer complementary routes to model robustness and selection~\cite{JMLR:v20:18-598,lu2020nsganetv2,duan2022ctnaseeg,lakshminarayanan2017simple}. In MI-EEG, however, the challenge is not only finding a high-validation model, but selecting a model or inference policy that transfers to a held-out subject. Table~\ref{tab:related_positioning} separates architecture compactness, subject-independent evaluation, automated selection, and validation control; \rev{among the representative families shown, ThinkNet is distinguished by making the source-validation-to-held-out selection gap an explicit analysis target.}

\begin{figure*}[t]
    \centering
\includegraphics[width=0.94\linewidth]{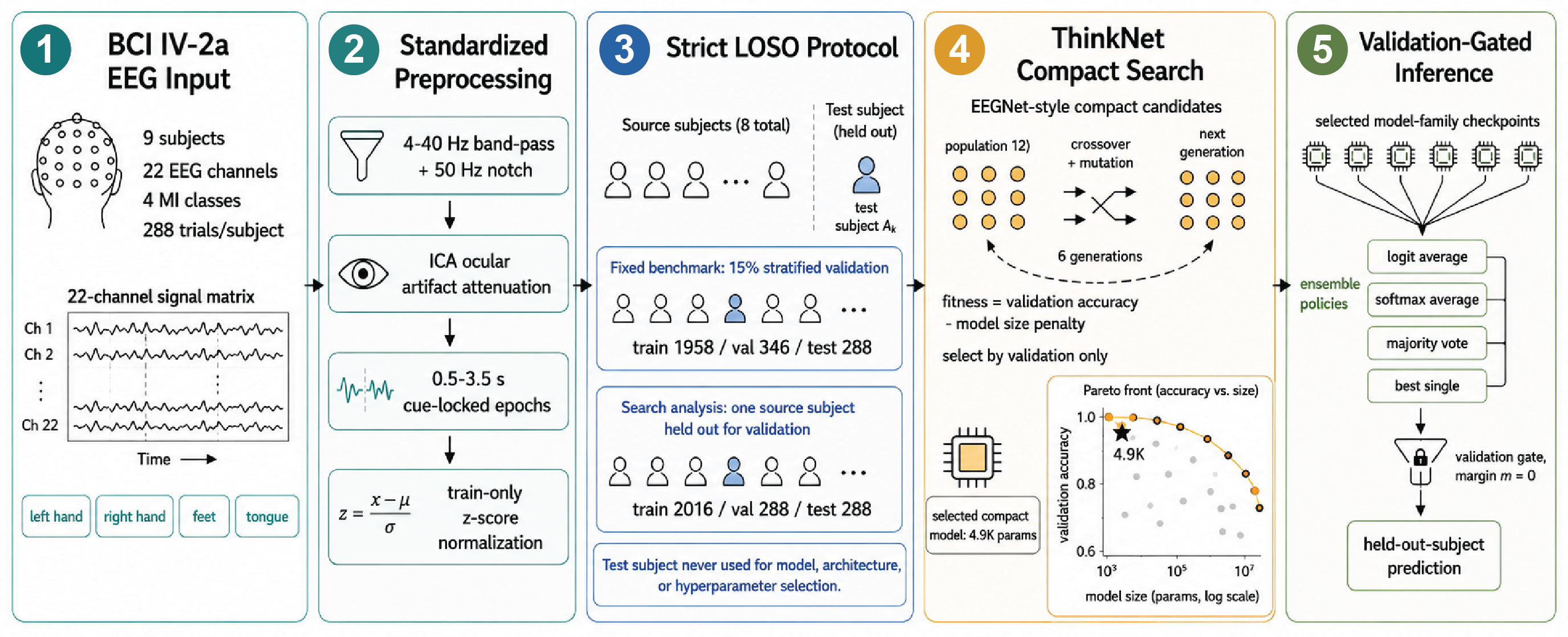}
   \caption{ThinkNet methodology. \rev{In fold-specific search and gating, source }data support normalization, validation, and selection \rev{without }held-out-subject feedback. Fixed unsupervised signal cleaning is applied without labels or model feedback.}
   \label{fig:methodology}
\end{figure*}

\section{ThinkNet Methodology}
\label{sec:methods}

\subsection{Overview}
ThinkNet is an accuracy--efficiency framework for subject-independent MI-EEG decoding. Its stages are preprocessing, LOSO splitting, evolutionary search, selected retraining, and validation-gated inference (Fig.~\ref{fig:methodology}). \rev{In fold-specific search and gating, no }held-out-subject labels, metrics, normalization statistics, or selection feedback are used for model selection, ensemble selection, early stopping, or hyperparameter choice; \rev{the fixed benchmark }is treated separately \rev{in Sec.~III-F. }\rev{This is a zero-target selection setting: target adaptation using unlabeled or labeled target epochs is not performed.}
\subsection{Problem Definition}
Given labeled EEG trials $\mathcal{D}=\{(\mathbf{X}_i,y_i,s_i)\}_{i=1}^{N}$, let $\mathbf{X}_i\in\mathbb{R}^{C\times T}$ denote a $C$-channel, $T$-sample epoch, $y_i\in\mathcal{Y}=\{1,\ldots,K\}$ the class label ($K=4$ here), and $s_i\in\mathcal{S}$ the subject identity. In a LOSO fold, one subject $s^{\ast}$ is held out for testing. A decoder $f_\theta$ with configuration $\theta\in\Theta$ is trained on source subjects and selected using source-validation data that excludes $s^{\ast}$:
\begin{equation}
    \theta^{\ast}=\arg\max_{\theta\in\Theta} \; \mathrm{Acc}_{\mathrm{val}}(f_\theta), \quad
    \mathrm{reported}=\mathrm{Acc}_{\mathrm{test}}(f_{\theta^{\ast}}).
    \label{eq:val_selection}
\end{equation}
Here $\Theta$ denotes the candidate configuration set. $\mathrm{Acc}_{\mathrm{val}}$ is computed on source-validation trials, whereas $\mathrm{Acc}_{\mathrm{test}}$ is computed on the held-out subject only after $\theta^{\ast}$ has been fixed. For search experiments, validation selection uses the accuracy--size fitness
\begin{equation}
    F(\theta)=w_a\,\widetilde{\mathrm{Acc}_{\mathrm{val}}}(\theta)
    -w_p\,\widetilde{\log(\mathrm{Params})}(\theta),
\label{eq:validation_fitness}
\end{equation}
where $\mathrm{Params}$ is trainable parameter count, tildes denote min--max normalization within the evaluated population, and $w_a,w_p\geq0$ control the accuracy--size trade-off. Eq.~\eqref{eq:val_selection} describes generic validation-controlled selection; in evolutionary search, $\mathrm{Acc}_{\mathrm{val}}$ is replaced by $F(\theta)$ in Eq.~\eqref{eq:validation_fitness}. The compact \rev{decoder} search uses $(w_a,w_p)=(0.7,0.3)$, while the \rev{broader family} search uses $(0.75,0.25)$; both are fixed before held-out testing. The held-out subject is never used to compute validation accuracy, fitness, or fitness-normalization terms.

\subsection{Preprocessing}
\begin{figure}[ht]
    \centering
\includegraphics[width=0.80\linewidth]{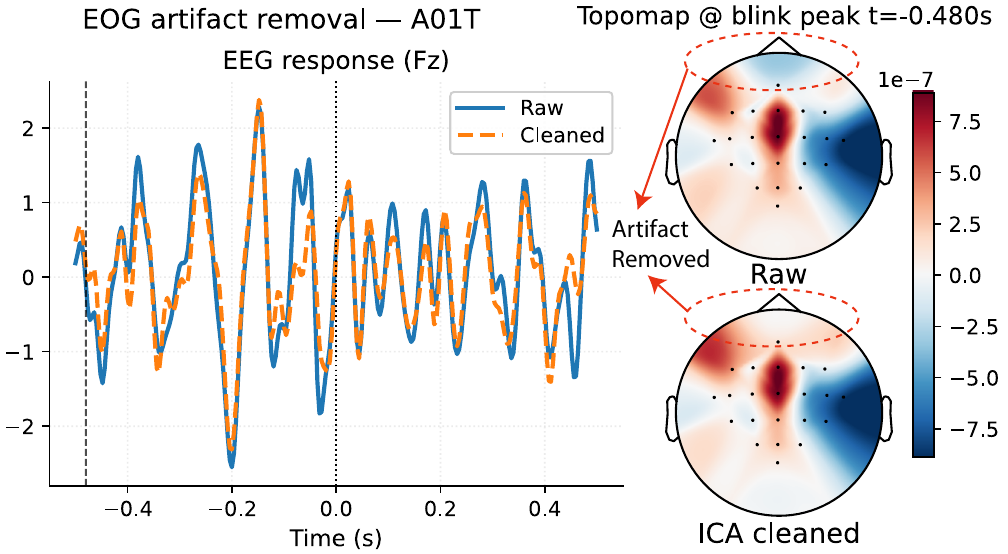}
    \caption{Example fixed, unsupervised ICA-based ocular artifact attenuation on BCI IV-2a subject A01T. The key observation is that ocular cleaning is performed before LOSO training without class labels, model outputs, or selection feedback.}
    \label{fig:eog_artifact}
\end{figure}

All models use a common preprocessing protocol to avoid architecture-specific signal conditioning. Raw trials are repaired for missing samples, cleaned with EOG-guided ICA for ocular attenuation, band-pass filtered from 4--40 Hz with a 50 Hz notch filter, average referenced, and epoched from 0.5--3.5 s after the cue. \rev{ICA is independently estimated within each recording as }unsupervised signal cleaning \rev{(Fig.~\ref{fig:eog_artifact}). Component }selection uses EOG correlation and is not informed by \rev{labels, }predictions, validation \rev{scores, or selection feedback. The only normalization statistic transferred from source training to validation and test trials is the }channel-wise z-score\rev{, }estimated exclusively from source-training trials \rev{and then applied unchanged}. Architectures requiring 128 Hz input use \rev{fixed resampling}, while the temporal window, labels, LOSO splits, and normalization rule remain identical across models.

\subsection{Evolutionary Configuration Search}
ThinkNet uses evolutionary search as a validation-controlled model-selection mechanism, not as a search tied to one architecture. The compact search tests whether a low-parameter \rev{depthwise-separable decoder family offers }a favorable accuracy--efficiency trade-off under subject shift. Each individual specifies \rev{filters, depth}, kernel width, pooling, dropout, optimizer, learning rate, weight decay, batch size, and training budget. Selection and fitness use source-subject validation only\rev{. Each fold uses one fixed-map }source subject for validation \rev{and seven }for training, \rev{with }population 12, six generations, three seeds, \rev{elite reuse, and duplicate skipping. This yields }61 candidates per fold-seed and 1647 overall.

The broader search applies the same principle across ten representative \rev{families spanning }compact EEG-specific CNNs, \rev{temporal convolution, filter banks}, attention/\rev{conformers}, and larger convolutional baselines. \rev{Ten is the benchmark set, not a method requirement. }Evolutionary search handles mixed categorical, discrete, and conditional choices. \rev{Each }fold-family task \rev{uses }24 candidates\rev{, eight generations, }three elites, tournament size 3, crossover 0.7, and mutation 0.3, yielding \rev{171 candidates}. Across nine folds and ten families, this gives 15,390 \rev{trainings; selected configurations are retrained }with seeds 42--44. %
\rev{Architecture and training choices are searched and logged. The matched random control runs one 171-candidate realization per fold-family (seed 42 with deterministic fold/ID/family offsets), then retrains seeds 42--44. Population-normalized fitness values are not compared across strategies.}

\subsection{\rev{Selector Diagnostics}}
\rev{For family set $\mathcal{M}$, we compare validation selection $a_{j^*}^{\mathrm{test}}$, $j^*=\arg\max_j a_j^{\mathrm{val}}$, with uniform-choice expectation $|\mathcal{M}|^{-1}\sum_j a_j^{\mathrm{test}}$ and the non-deployable test-best reference $\max_j a_j^{\mathrm{test}}$; selector regret is test-best minus selected accuracy. A fixed-family control selects EEGNet configurations source-only within fold and differs from Table~\ref{tab:main_loso_results}. Rank transfer is within-fold Spearman correlation after seed averaging. Confidence intervals bootstrap subjects and are descriptive because LOSO source-training sets overlap.}

\rev{A validation-coverage sensitivity independently retrains all ten selected family configurations with seeds 42--44. The one-source arm uses the next cyclic source subject. The three-source arm pools trials from the next three cyclic source subjects into one validation set. The remaining seven or five subjects form the corresponding training set. Family scores are accuracies over the pooled validation trials, and the selector and rank diagnostics above are applied unchanged.}
\subsection{Fixed Baselines and Validation-Gated Ensembles}
The fixed benchmark \rev{is a frozen post-discovery comparison. The compact ThinkNet}, CNN, and \rev{Transformer entries were selected by validation fitness using A01--A07 for training and A08 for validation; A09 did not influence configuration choice. The entries were }then frozen before \rev{nine-fold }LOSO retraining. \rev{Because discovery was not nested by fold, central target-isolated claims rest on fold-specific search and gating. Reference EEGNet }is a fixed compact control; EEG-TCNet, Shallow ConvNet, and EEG-Inception use reference-style settings. For ensemble analysis, one validation-selected checkpoint per family feeds \rev{best-single, logit-average, softmax-average, and majority-vote policies}. The compact six-family \rev{set is fixed }by a parameter-only screen (median selected size $\leq$25K), \rev{without validation or test accuracy }for membership.
\begin{figure}[ht]
    \centering
\includegraphics[width=0.94\linewidth]{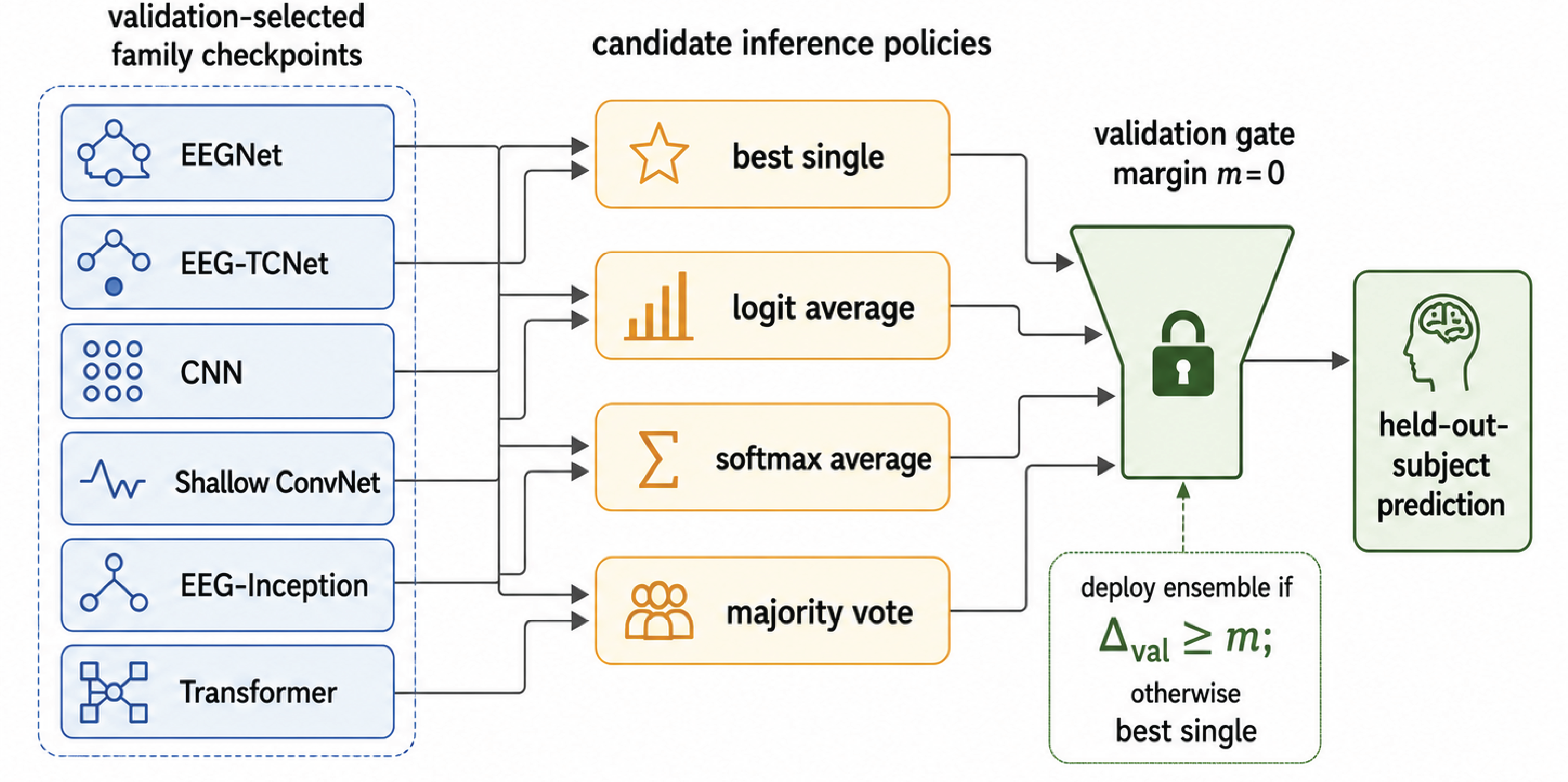}
    \caption{Validation-gated ensemble selection. \rev{Validation }logits choose the policy; \rev{held-out }logits are applied only after the gate is fixed.}
    \label{fig:gated_ensemble_protocol}
\end{figure}

The deployed policy is chosen by a validation gate. Let $\pi_{\mathrm{single}}$ be the validation-best individual model and $\pi_{\mathrm{ens}}^{\ast}$ the validation-best ensemble among raw-logit averaging, softmax averaging, and majority voting. Majority-vote ties resolve to the lowest class index. With $\Delta_{\mathrm{val}}=\mathrm{Acc}_{\mathrm{val}}(\pi_{\mathrm{ens}}^{\ast})-\mathrm{Acc}_{\mathrm{val}}(\pi_{\mathrm{single}})$,
\begin{equation}
    \pi^{\ast}=\begin{cases}
    \pi_{\mathrm{ens}}^{\ast}, & \Delta_{\mathrm{val}}\geq m,\\
    \pi_{\mathrm{single}}, & \Delta_{\mathrm{val}}<m.
    \end{cases}
\end{equation}
We use a pre-specified margin $m=0$ in all reported ensemble experiments; validation ties follow the pre-specified candidate-policy order, with no test feedback (Fig.~\ref{fig:gated_ensemble_protocol}).

In Algorithm~\ref{algo:gated_ensemble}, $\mathcal{Z}^{\mathrm{val}}$ and $\mathcal{Z}^{\mathrm{test}}$ are precomputed logits from $M$ checkpoints, $\Pi_{\mathrm{single}}$ and $\Pi_{\mathrm{ens}}$ are policy sets, and scoring $P$ ensemble policies over $N_{\mathrm{val}}$ validation trials scales as $O(PMN_{\mathrm{val}}K)$; here $P=3$ and $K=4$.

\begin{algorithm}
\caption{Validation-Gated Ensemble Selection}\label{algo:gated_ensemble}
\footnotesize
\begin{algorithmic}[1]
\STATE \textbf{Input:} validation logits $\mathcal{Z}^{\mathrm{val}}$, held-out logits $\mathcal{Z}^{\mathrm{test}}$, margin $m$ \quad \textbf{Output:} selected policy $\pi^{\ast}$ and held-out prediction $\hat{y}^{\mathrm{test}}$
\STATE $\Pi_{\mathrm{single}}\leftarrow$ individual checkpoints;
\STATE $\Pi_{\mathrm{ens}}\leftarrow\{\text{logit avg.},\text{softmax avg.},\text{majority vote}\}$
\STATE $\pi_{\mathrm{single}}\leftarrow\arg\max_{\pi\in\Pi_{\mathrm{single}}}\mathrm{Acc}_{\mathrm{val}}(\pi)$
\STATE $\pi_{\mathrm{ens}}^{\ast}\leftarrow\arg\max_{\pi\in\Pi_{\mathrm{ens}}}\mathrm{Acc}_{\mathrm{val}}(\pi)$ \hfill \textit{// validation only}
\STATE $\Delta_{\mathrm{val}}\leftarrow\mathrm{Acc}_{\mathrm{val}}(\pi_{\mathrm{ens}}^{\ast})-\mathrm{Acc}_{\mathrm{val}}(\pi_{\mathrm{single}})$
\IF{$\Delta_{\mathrm{val}}\geq m$}
    \STATE \hspace{0.8em}$\pi^{\ast}\leftarrow\pi_{\mathrm{ens}}^{\ast}$
\ELSE
    \STATE \hspace{0.8em}$\pi^{\ast}\leftarrow\pi_{\mathrm{single}}$
\ENDIF
\STATE $\hat{y}^{\mathrm{test}}\leftarrow\pi^{\ast}(\mathcal{Z}^{\mathrm{test}})$ \hfill \textit{// policy fixed before test scoring}
\RETURN $\pi^{\ast}$, $\hat{y}^{\mathrm{test}}$, and reported held-out-subject accuracy
\end{algorithmic}
\end{algorithm}

\section{Experimental Setup}
\label{sec:experiments}
\subsection{Dataset and LOSO Protocol}
We evaluate on BCI Competition IV-2a~\cite{tangermann2012review}: nine subjects (A01--A09), 22 EEG channels, four motor-imagery classes, and 288 trials/session. EOG channels are used only for artifact scoring and not as decoder inputs. All experiments use session T as a fixed single-session LOSO protocol\rev{; cross-session }generalization is outside scope. Each fold holds out one subject ($n=288$), leaving eight source subjects. The fixed comparison uses stratified 15\% source-trial validation ($n=346$; train $n=1958$)\rev{, whereas }search uses one \rev{validation subject }($n=288$) and seven \rev{training subjects }($n=2016$). \rev{Compact }search uses a pre-specified seed-dependent \rev{validation map. Broader search uses the next cyclic subject}, fixed across families and candidates\rev{. }\rev{The Lee2019/OpenBMI session-1 check~\mbox{%
\cite{lee2019openbmi} }\hskip0pt%
has 54 subjects, 62 channels, two classes, and 100 trials/subject. Before inspecting its outcomes, we fixed five submitted configurations representing compact/reference EEG, temporal-convolution, large-CNN, and Transformer categories that required no architecture redesign beyond dataset-determined input/output dimensions. Each fold uses 52 training, one cyclic validation, and one test subject; search is not repeated.}
\subsection{Models and Training}
The \rev{frozen post-discovery benchmark evaluates the }compact decoder, reference EEGNet, EEG-TCNet, Shallow ConvNet, EEG-Inception, CNN, and Transformer. \rev{Discovery-selected entries are compact }ID 42 ($F_1{=}32$, $D{=}1$, kernel 32), CNN ID 129 (five depthwise temporal layers), and Transformer ID 69 (conv front-end, $d{=}128$, 16 heads, four layers). Reference-style entries use \rev{recorded budgets }(500--750 \rev{epochs; }patience 50--75). All models use cross-entropy, validation-accuracy early stopping, shuffled \rev{source }loaders, and no \rev{target }adaptation, test-time augmentation, class weighting, or learning-rate scheduling. The \rev{benchmark has }189 runs ($9\times3\times7$).

\subsection{Search and Ensemble Evaluation}
The compact search evaluates 27 fold-seed runs and 1647 candidates. The broader search evaluates 90 fold-family tasks and 15,390 \rev{trainings, then }270 selected-configuration \rev{retrainings }($9\times10\times3$\rev{; }seeds 42--44\rev{). }\rev{Matched random search uses the same spaces, 171 evaluations/task, budget, fitness, and retraining. The independently retrained validation-coverage sensitivity uses all ten families and the same three seeds. Moving from one to three pooled validation subjects reduces training subjects from seven to five, so its paired results are descriptive rather than causal.} \rev{The }non-deployable \rev{test-best reference }selects the highest \rev{test accuracy only to quantify selector regret. Validation }logits choose the gated policy \rev{(}$m=0$\rev{) before }held-out logits are scored.

\begin{figure*}[b]
    \centering
\includegraphics[width=0.945\linewidth]{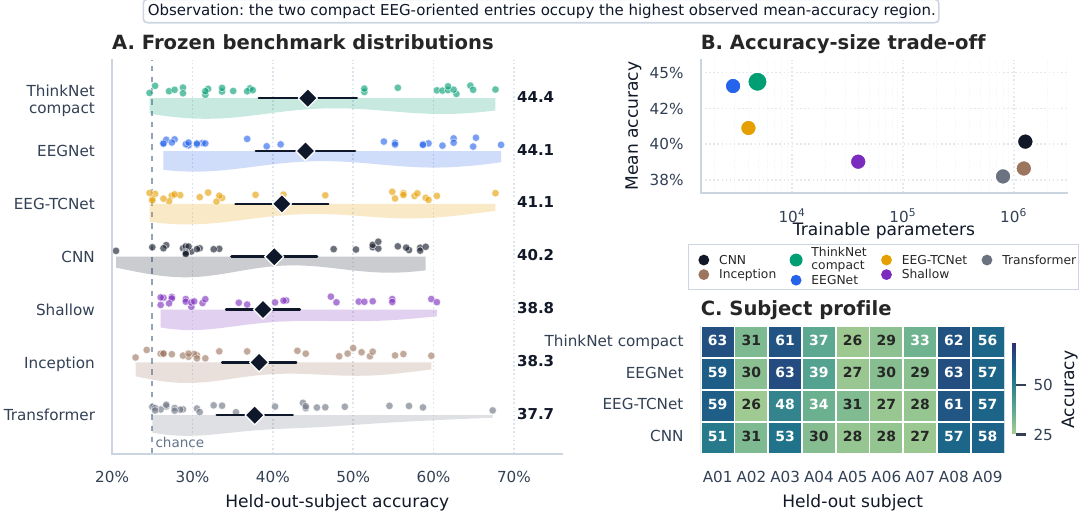}
    \caption{(A) Fold-seed distributions show large subject-shift variation. (B) Accuracy--size trade-off shows compact models in the high-accuracy region. (C) Subject profiles show that several held-out subjects remain difficult across decoder families.}
    \label{fig:main_results_compact}
\end{figure*}

\subsection{Metrics and Reproducibility}
The primary metric is held-out-subject accuracy\rev{; we also report parameters and subject profiles. }\rev{Fixed-benchmark means and SDs summarize nine seed-averaged subject scores. Other controls likewise average seeds within subject before 95\% bootstrap intervals over nine BCI or 54 Lee2019 subjects. These intervals are descriptive because LOSO source-training sets overlap.} FP32 memory \rev{uses }trainable weights. Orin \rev{latency is }batch-1 \rev{FP32 forward time on }model-specific \rev{synthetic input }after 100 \rev{warm-ups }and five 500-iteration \rev{repeats }with CUDA synchronization\rev{. }\rev{Measurements used a Jetson AGX Orin 64 GB in MAXN mode (Jetson Linux R36.4.4, CUDA 12.6, PyTorch 2.7.0); ensemble timing includes serial member forwards and aggregation, excluding acquisition/preprocessing.} \rev{Stored splits, configurations}, checkpoints, and \rev{CSVs trace each aggregate to }fold, seed, family, and checkpoint. %
\rev{The bundle retains the submitted fixed/search/retraining records and adds matched-random-search, Lee2019, validation-coverage sensitivity, subject-bootstrap, and Orin timing artifacts.}
 
\section{Results, Analysis, and Discussion}
\label{sec:results}

\begin{figure*}[ht]
    \centering
\includegraphics[width=0.87\linewidth]{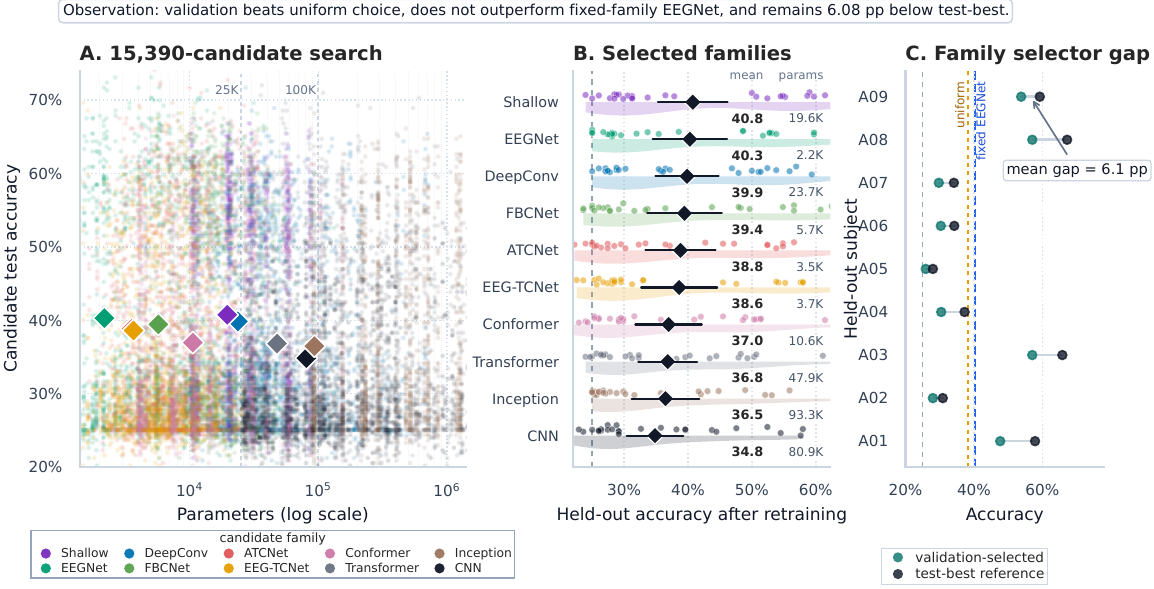}
    \caption{\rev{Search }and selection reliability. (A) Candidate landscape; test accuracy is post hoc and never used for selection. (B) Selected-family distributions. (C) \rev{Seed-averaged subject results for }validation-selected versus non-deployable \rev{test-best }family accuracy\rev{; vertical lines mark uniform-choice expectation and fixed-family EEGNet}.}
    \label{fig:search_landscape}
\end{figure*}

\subsection{\rev{Selection Reliability and Search Controls}}
\rev{\textbf{Selector reliability:} Source-validation selection reached 40.01\%, exceeding the uniform-choice expectation by 1.72 points (95\% CI $[0.52,3.13]$), but it did not outperform fixed-family EEGNet (40.29\%; difference $-0.28$, $[-2.01,1.32]$). The non-deployable test-best family reached 46.09\%, leaving 6.08 points of selector regret. Within-fold rank transfer was weak ($\rho=0.256$, $[0.04,0.46]$; pooled $\rho=0.021$). Thus, target isolation prevents selection leakage, but source validation does not reliably identify the best family for an unseen subject.
}

\rev{\textbf{Compactness:} Fig.~\ref{fig:search_landscape} places these references beside the full search log. Restricting eligibility using the pre-specified $\leq25$K compact threshold changed selected accuracy by only $-0.05$ points (39.96\% vs. 40.01\%; $[-0.54,0.39]$), and compact entries averaged 2.83 points above noncompact entries ($[1.24,4.63]$). After retraining, the compact, mid-size, and large groups averaged 40.10$\pm$14.06\%, 35.09$\pm$11.52\%, and 34.78$\pm$11.32\%, respectively. Separately, the compact candidate search exhibited a 7.90-point candidate-level diagnostic gap; this is not the primary family-level regret above.
}

\rev{\textbf{Search control:} In this single-trajectory-per-task, budget-matched control, evolution had higher observed source-validation accuracy (52.71\% vs. 51.46\%) and a 22.8\% lower geometric-mean selected parameter count (ratio 0.77, $[0.60,0.99]$), without detected held-out gain ($-0.81$ points, $[-1.87,0.15]$). Thus, its observed benefit is smaller configurations rather than unseen-subject accuracy. Increasing validation subjects from one to three changed $\rho$ by 0.126 ($[-0.014,0.283]$), regret by $-1.49$ points ($[-4.33,2.29]$), and accuracy by +0.37 points ($[-4.50,4.54]$). Because training subjects simultaneously fell from seven to five, these sensitivity trends are descriptive.
}

\begin{wraptable}[7]{r}{0.45\columnwidth}
\vspace{-20pt}
\begingroup
\centering
\caption{\rev{Independent Lee2019 check.}}
\label{tab:lee_control}
\scriptsize
\setlength{\tabcolsep}{1.2pt}
\renewcommand{\arraystretch}{0.84}
\resizebox{\linewidth}{!}{%
\begin{tabular}{@{}lrr@{}}
\toprule
\textbf{\rev{Frozen model}} & \textbf{\rev{Params}} & \textbf{\rev{Acc. (\%)}} \\
\midrule
\rev{Reference EEGNet }& \rev{2.8K }& \rev{67.33 }\\
\rev{Submitted compact }& \rev{5.4K }& \rev{65.22 }\\
\rev{EEG-TCNet }& \rev{4.7K }& \rev{65.44 }\\
\rev{CNN }& \rev{1.26M }& \rev{61.54 }\\
\rev{Transformer }& \rev{804K }& \rev{64.57 }\\
\bottomrule
\end{tabular}}
\endgroup
\vspace{-6pt}
\end{wraptable}
{ \rev{Among five frozen configurations on Lee2019 session 1, compact entries averaged 2.94 points above larger entries ($[0.65,5.34]$). The selector hierarchy was 64.82\% uniform choice, 66.63\% validation selection, 67.33\% reference EEGNet, and 71.94\% test-best; within-fold $\rho=0.281$ ($[0.13,0.42]$). Because Lee2019 is two-class and BCI IV-2a is four-class, this independent check concerns relative size--performance and selector-regret behavior, not absolute accuracy, repeated search, or cross-session transfer. Table~\ref{tab:lee_control} summarizes this independent Lee2019 check across the five frozen configurations.
}\par}
\WFclear

\subsection{\rev{Frozen Post-Discovery}\ Benchmark}
Table~\ref{tab:main_loso_results} reports the \rev{frozen post-discovery }benchmark and resource accounting. The \rev{selected compact decoder reached }44.35$\pm$\rev{15.70\% with }4.9K parameters, 19 \rev{KiB\ }FP32 weights, and 0.99 ms Orin CUDA latency. \rev{Its paired subject-level difference from reference EEGNet was +0.30 points (95\% bootstrap CI $[-1.07,1.84]$), so no accuracy advantage is claimed; other entries had observed means }3.23--6.61 \rev{points lower. The engineering observation is that the two compact EEG-oriented entries occupy the highest observed mean-accuracy region, while subject identity remains a major source of variation (Fig.~\ref{fig:main_results_compact}}).

\vspace{-5pt}
\begin{table}[ht]
\centering
\caption{Benchmark: \rev{mean$\pm$SD over nine seed-averaged subjects}.}
\label{tab:main_loso_results}
\scriptsize
\begin{tabular}{lcccc}
\toprule
\textbf{Model} & \textbf{Params} & \textbf{Mem.} & \textbf{Orin ms} & \textbf{Test Acc. (\%)} \\
\midrule
ThinkNet compact & 4.9K & 19 \rev{KiB}& 0.99 & \textbf{44.35$\pm$\rev{15.70}} \\
EEGNet baseline & 2.9K & 11 \rev{KiB}& 0.98 & \textbf{44.06$\pm$\rev{16.04}} \\
EEG-TCNet & 4.0K & 16 \rev{KiB}& 2.73 & 41.13$\pm$\rev{14.81 }\\
CNN & 1.27M & 4.9 \rev{MiB}& 1.84 & 40.17$\pm$\rev{13.78 }\\
Shallow ConvNet & 39.5K & 154 \rev{KiB}& 0.56 & 38.77$\pm$\rev{11.89 }\\
EEG-Inception & 1.23M & 4.8 \rev{MiB}& 7.23 & 38.30$\pm$\rev{11.61 }\\
Transformer & 800K & 3.1 \rev{MiB}& 3.86 & 37.74$\pm$\rev{12.06 }\\
\bottomrule
\end{tabular}
\end{table}
\begin{figure*}[!t]
    \centering
\includegraphics[width=0.90\linewidth]{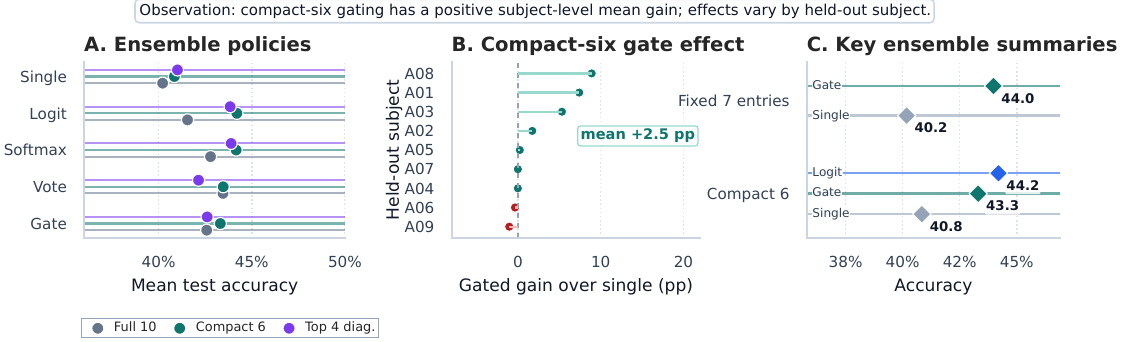}
    \caption{Validation-gated analysis. (A) \rev{Policy means and subject-level }95\% CIs. (B) \rev{Compact-six gains after averaging seeds within each held-out subject}. (C) \rev{Summary. The positive mean gain varies by subject}.}
    \label{fig:ensemble_selector}
\end{figure*}
\subsection{Validation-Gated Ensemble}
Validation-gated \rev{ensembling increased the observed mean by }3.81 \rev{points in the fixed benchmark and from }40.84\% to 43.31\rev{\% for compact-six. After subject-level seed averaging, compact-six gained 2.47 points ($[0.39,4.89]$): 5/9 subjects improved, two tied, and two worsened. Lee2019 gains were 0.78 points for compact-three ($[-0.33,1.94]$) and 1.85 for all five ($[0.72,3.07]$). These are candidate-set-dependent mean gains, not uniform improvements or evidence of a causal diversity effect.
}\begin{wraptable}[7]{r}{0.55\columnwidth}
\vspace{-15pt}
\begingroup
\centering
\caption{\rev{Serial Orin FP32 cost (batch 1).}}
\label{tab:orin_control}
\scriptsize
\setlength{\tabcolsep}{1.2pt}
\renewcommand{\arraystretch}{0.84}
\resizebox{\linewidth}{!}{%
\begin{tabular}{@{}lrrr@{}}
\toprule
\textbf{\rev{Operating point}} & \textbf{\rev{Params}} & \textbf{\rev{MiB}} & \textbf{\rev{ms}} \\
\midrule
\rev{Validation-best individual }& \rev{18.0K }& \rev{.069 }& \rev{2.24 }\\
\rev{Compact-six logit avg. }& \rev{64.1K }& \rev{.244 }& \rev{9.41 }\\
\rev{Full-ten logit avg. }& \rev{464.9K }& \rev{1.773 }& \rev{20.98 }\\
\rev{Full-ten realized gate }& \rev{-- }& \rev{-- }& \rev{12.15 }\\
\bottomrule
\end{tabular}}
\endgroup
\vspace{-6pt}
\end{wraptable}
{\rev{Measurements include serial member forwards and aggregation. Compact-six selected an ensemble in 13/27 groups and a single model in 14/27; measured ensemble branches required 9.41--9.49 ms. The full-ten gate selected a single in 13/27 and an ensemble in 14/27, yielding a 12.15-ms realized mean. Gating therefore provides an optional accuracy--cost operating point; mean benefits vary by subject and incur additional inference cost. Table~\ref{tab:orin_control} summarizes the measured Orin cost of the different operating points.}\par}
\WFclear

\subsection{Discussion}
\rev{Three conclusions follow}. First, \rev{target isolation is necessary for valid evaluation but does not ensure reliable family selection: validation beats uniform choice, not fixed-family EEGNet, and leaves substantial regret}. Second, \rev{restricting eligibility to the compact set produced little observed mean change in this benchmark, while matched random search limits evolution's supported role to compact mixed-space optimization. If one strong compact default is sufficient, full search is not necessary relative to fixed-family EEGNet. Its value here is auditing family-selection reliability and optimizing mixed accuracy--size spaces}. Third, \rev{validation-gated policies show positive mean gains for some candidate sets, but benefits vary by subject and incur measured inference cost. ThinkNet therefore separates whether a fixed compact default is sufficient from whether validation evidence supports a different model or higher-cost policy. It is }not a universal state-of-the-art \rev{or clinical claim; online, cross-session, physiological-causal, and clinical effectiveness remain outside the evidence}.

\section{Conclusion}
\label{sec:conclusion}
ThinkNet provides a reproducible, validation-controlled framework for compact, subject-independent MI-EEG decoding under strict LOSO evaluation. \rev{Its central result is that source validation is measurably informative relative to uniform choice but did not outperform fixed-family EEGNet, showed weak within-fold rank transfer, and remained 6.08 points below the test-best family. In a single-trajectory-per-task matched control, evolution selected smaller models without detected held-out gain, so the broad search is best viewed as a selector diagnostic and compactness optimizer. Among five frozen configurations, Lee2019 provided an independent consistency check for the size--performance and selector-regret patterns. Separately, the post-discovery compact decoder achieved }44.35$\pm$\rev{15.70\% with }4.9K parameters and 0.99 ms \rev{Orin inference. Gated ensembles offered candidate-set-dependent mean gains at higher measured cost. Because the evidence is offline and }single-session\rev{, it does not establish }universal state-of-the-art, cross-session\rev{, online}, or clinical \rev{effectiveness. Future }work should prioritize subject-shift-aware validation \rev{and target adaptation}.

\balance
\begingroup
\makeatletter
\let\oldthebibliography\thebibliography
\let\endoldthebibliography\endthebibliography
\renewenvironment{thebibliography}[1]{%
  \oldthebibliography{#1}%
  \renewcommand{\baselinestretch}{0.86}\selectfont%
  \setlength{\itemsep}{-1pt}%
  \setlength{\parsep}{0pt}%
  \setlength{\parskip}{0pt}%
  \let\originalbibitem\bibitem
  \renewcommand{\bibitem}[1]{%
    \color{black}%
    \def\currentbibkey{##1}%
    \def\revisedbibkey{lee2019openbmi}%
    \ifx\currentbibkey\revisedbibkey\color{black}\fi
    \originalbibitem{##1}%
  }%
}{\endoldthebibliography}
\makeatother
\bibliographystyle{ieeetr}
\bibliography{cite}
\endgroup

\end{document}